\documentclass[seceq]{jpsj2}
\title{Satellite Holmium $M$-Edge Spectra 
from the Magnetic Phase \\
via Resonant Soft X-Ray Scattering}

\author{Tatsuya \textsc{Nagao}$^{1}$\thanks{E-mail:tnagao@phys.sci.gunma-u.ac.jp} and Jun-ichi \textsc{Igarashi}$^{2}$}
\inst{$^{1}$Faculty of Engineering, Gunma University, Kiryu, Gunma 376-8515, Japan \\
$^{2}$Faculty of Science, Ibaraki University, Mito, Ibaraki 310-8512, Japan
}
\abst{Developing an expression of resonant x-ray scattering (RXS) amplitude
which is convenient for investigating the contributions
from the higher rank tensor on the basis of a localized electron picture,
we analyze the RXS spectra from the magnetic phases of Ho near the
$M_{4,5}$ absorption edges.
At the $M_5$ edge in the uniform helical
phase, the calculated spectra of the absorption coefficient,
the RXS intensities at the first and second satellite spots
capture the properties the experimental data possess,
such as 
the spectral shapes and the peak positions.
This demonstrates the plausibility of the adoption of the
localized picture in this material and
the effectiveness of the spectral shape analysis.
The latter point is markedly valuable since the azimuthal
angle dependence, which is one of the most useful informations
RXS can provides, is lacking in the experimental conditions.
Then, by focusing on the temperature dependence of the
spectral shape at the second satellite spot,
we expect that the spectrum is the contribution
of the pure rank two profile in the uniform helical
and the conical phases while that is dominated by the
rank one profile in the intermediate temperature phase,
so-called spin slip phase.
The change of the spectral shape as a function of
temperature indicates a direct evidence
of the change of magnetic structures undergoing. 
Furthermore, we predict that the intensity, which is
the same order observed at the second satellite spot,
is expected at the fourth satellite spot from the conical
phase in the
electric dipolar transition.}
\kword{resonant x-ray scattering, Ho metal, 
amplitude formula, satellite spot signal,
helical magnetic structure}

\begin{document}
\maketitle

\section{\label{sect.1}Introduction}

With the advent of the so-called third generation
synchrotron sources, resonant x-ray scattering
(RXS) has opened a new research field  of investigating,
for example, the multipole ordering phases in the
$f$ electron systems due to the electric multipolar transitions.
RXS is a second order optical process that an incident
photon with the energy tuned through an absorption edge
excites the core electron to an unoccupied level, 
then the excited electron decays into the core
level accompanied by emission of a photon.
Element, electron shell, and chemical valence specific 
natures of the RXS technique
have made it possible to become a complement to
neutron scattering to elucidate the spatial Fourier
transform of the multipole order such as
quadrupole and octupole as well as dipole 
moments.\cite{Murakami98,Nakao01,Yakhou01,Tanaka99,
Hirota00,Paixao02,Mannix05}

Historically, however, analysis of the magnetic 
properties on the magnetic materials
in terms of the x-ray resonant process, so-called 
resonant magnetic x-ray scattering (RMXS), 
developed in its own right
since Blume suggested a possibility of such 
a usage.\cite{Blume85}
The first observation of RMXS was made by 
Namikawa \textit{et al}. on ferromagnetic 
nickel.\cite{Namikawa85} 
Since then, further studies followed,
investigating, in particular, the $f$ electron systems
such as holmium, uranium 
arsenide, and so on.\cite{Gibbs88,McWhan90,Tang92}

Metallic holmium is one of the most studied materials
by using RMXS. 
The crystal structure of holmium is hexagonal close packed 
structure with lattice constants $a=3.58$ $\textrm{\AA}$
and $c=5.62$ $\textrm{\AA}$ at room temperature.
On cooling down the temperature, magnetic moments
in Ho order at $T_{\textrm{N}}=133$ K 
with a helical magnetic structure,
which consists of the ferromagnetically aligned
moments within each basal plane and the orientation
of the moments rotates in successive basal planes. 
As the temperature decreases, the wavevector
characterizing the spatial modulation
of the helical pattern reduces and locks in
below $T_{\textrm{C}}=20$ K where
the magnetic structure
becomes a conical configuration.\cite{Koehler66,Koehler67}

Gibbs \textit{et al}. detected the RMXS signals
near the Ho $L_3$ edge in the magnetic satellite
spots.\cite{Gibbs88}
Polarization analysis distinguished two kinds of spectra
with peak positions separated about 6 eV.
They are attributed to the consequences
of the electric dipole ($E$1) 
and quadrupole ($E$2) transitions.
The former had intensities up to the second 
harmonic satellite spot 
while the latter was up to the fourth
harmonic satellite spot.
Hannon \textit{et al}. gave the theoretical
explanation of the experimental results.\cite{Hannon88}
They derived useful formulae describing the
RMXS amplitude and, by using them,
concluded that relatively large scattering amplitude
would be expected in the vicinity of
the Ho $L_{2,3}$ and $M_{4,5}$
edges, the former of
which explained the experiment. 
By extending the formula derived by Hannon \textit{et al}.,
Hill and McMorrow investigated some qualitative
properties of the RXS signals at the Ho $L_3$ edge in the
uniform helimagnetic and the conical phases.\cite{Hill96}
These works have stimulated the immense activities in the
field of RMXS both experimentally and 
theoretically.\cite{Vettier01}

As for Ho, several experiments followed focusing,
for instance, on the temperature dependence of the
exponents, the existence of two-length scales, 
and so on.\cite{Gibbs91,Thurston94,Helgesen94}
Theoretical analyses based on the mean field model
revealed the importance of the interaction 
having the trigonal symmetry.\cite{Simpson95,Jensen96}
Despite the numerous achievements provided by the RMXS
data at the Ho $L_3$ edge, it is difficult to extract the
quantitative interpretation since the
$E$1 process at the $L_3$ edge is the consequence of 
the transition between the $2p$ and $5d$ states
so that the relevancy of the $4f$ electrons is indirect.
The band nature of the $5d$ electrons also increases
the complication of the analysis.
Although the $E$2 signal, the transition between the $2p$ and
$4f$ states, presents a direct information of the $4f$
electrons, the interference with the $E$1 signal
complicates the quantitative interpretation.

A possibility has long been suggested
that the RMXS measurement at the Ho
$M_{4,5}$ edges by means of soft x-ray beam
may supply the useful insights to elucidate the 
natures of the magnetic structures.\cite{Hannon88}
The $E$1 processes at the $M_{4,5}$ edges
are brought about by 
the transition between the $3d$ and $4f$ states, which
provides a direct information from the $4f$ electrons.
Furthermore, long wavelength of the soft x-ray is
suitable for investigating the long period magnetic
structure of Ho.
Recently, Spencer \textit{et al}.
and Ott \textit{et al}. have succeeded in detecting 
the RMXS spectra of the first and second harmonics satellite
spots as well as the absorption coefficient 
spectrum at the Ho $M_5$ edge from the
helical magnetic phase.\cite{Spencer05,Ott06}
The energy profiles of the spectra of the first and second
satellite spots differ significantly
from each other.  They were attributed to 
the circular and linear
dichroic contributions, respectively.
In this case, it turned out that
the absence of the $E$2 contribution 
made the analysis rather simple.
These facts make a sharp contrast observed at the $L_3$ edge
where the difference of the spectral shapes 
at the different satellite spots were not 
so prominent.\cite{Gibbs88}

Usually, the azimuthal angle $(\psi$) dependence of the 
peak intensity of the spectrum offers one of the most
highlighted outcomes given by the R(M)XS experiment
in many cases.  
In holmium, however, the intensity shows no $\psi$ dependence
in the uniform helical and conical phases
with the choice of the scattering vector adopted in the
experiment.
Therefore, we should extract implications as many as
possible through the spectral analysis.
Then, a theoretical study may lend support to the
analysis of the R(M)XS spectra at the $M_{4,5}$ edges.

In this paper, therefore, we perform a theoretical
spectral analysis of RMXS at the Ho M$_{4,5}$ edge.
Based on a localized electron picture,
we calculate the spectra exploiting a
useful expression of the scattering amplitude
whose utility is examined in several $f$ electron
systems.\cite{Nagao05,Nagao05.2,Nagao06}
We investigate three spectra; the RMXS spectra at the
first and second satellite spots and the absorption
spectrum.
Their shapes at the $M_5$ edge are distinguishable one another
and capture the main features presented
by the experiment, spectral shape and the shift of the
peak position at the different rank satellite spots.
The ratio of the peak intensity
extracted from the calculated spectra also
is the same order as the experimental one.

Another result we present is a prediction that
the spectral shape varies as a function of temperature.
For instance, the spectrum at the second satellite
spot shows the same energy profiles both in the
helical and conical phases.
On the other hand, the profile becomes completely different
one in the temperature range between two phases,
known so-called as the spin slip phase.
Such an evolution of the spectral shape may
provide the direct evidence that the
change of the magnetic structure is undergoing.

Further inferred from our investigation is a possibility
of the observation of the RXS spectra at the higher-order
satellite spots. In the conical phase, the presence
of fifth- and seventh-order magnetic satellites in the
neutron scattering were observed.\cite{Koehler66}
This is attributed to the fact that
the turn angle of the magnetic moment between successive 
Ho planes is not a constant (see Fig. \ref{fig.phase} (b)). 
Similarly, we show that the RXS intensities are expected
at the higher-order satellite spots in this phase.
For example, in the $E$1 transition, fourth-, fifth-,
seventh-, and eighth-order satellites are possible
to be detected.
Our estimation reveals that the RXS spectrum
at the Ho $M_5$ fourth-order satellite spot
from the $\pi-\sigma'$ channel possesses
the same order of magnitude and the same energy profile
as that observed by Ott \textit{et al}.
at the second satellite spot. 

The E$1$ transition at $M_{4,5}$ edges of rare earth metals give rise to
an excitation of electron from the $3d$ levels to the $4f$ levels.
Since the associated photon energy is about $1\sim 2$ KeV, this process
is only useful to investigate the long-range orders with rather long periods
in comparison with the lattice constant. In this paper, taking up the
helical and conical magnetic orders in Ho metal, we analyze the RXS
spectra in detail

The organization of this manuscript is as
follows.
In Sect. \ref{sect.2}, we explain
the various magnetic structures exhibited by Ho. 
In Sec. \ref{sect.3}, we present a brief
summary of the theoretical framework of
R(M)XS. An expression of the R(M)XS amplitude
the present authors had derived\cite{Nagao05,Nagao06}
is introduced and is compared with the
previous similar 
results.\cite{Hannon88,Carra94,Blume94,Hill96,
Lovesey96,Ovchinnikova00}
In Sec. \ref{sect.4},
the initial and the intermediate states needed to calculate
the RXS spectra are defined
 in order to apply the theory
to the investigation of the RMXS spectra
of the satellite spots
in the vicinity of the Ho $M_5$ edge.
Qualitative features expected
from each phases are analyzed in Sec. \ref{sect.5}.
Numerically calculated spectra of
the absorption coefficient and RMXS
are shown and compared with the experiment in
Sec. \ref{sect.6}.
Main conclusions are summarized in Sect. \ref{sect.7}.

\section{\label{sect.2}Magnetic ordering phases of Ho metal}

\begin{figure}[h]
\begin{center}
\includegraphics[width=8.0cm]{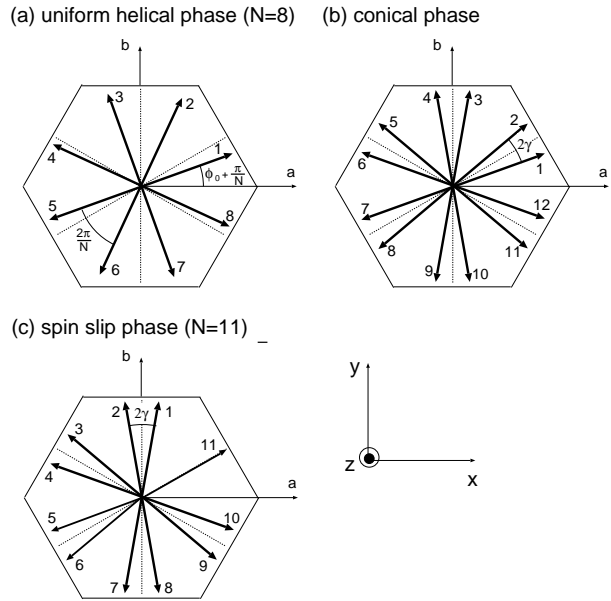}
\caption{Orientation of the magnetic
moment projected onto the $ab$ plane:
(a) uniform helical phase with $N=8$, (b)
conical phase, and (c) spin slip phase with $N=11$.
Direction of arrow associated with number $n=1, \cdots , N$
represents the direction of the magnetic
moment in the $n$-th Ho layer.
The dotted lines denote six equivalent easy axes.
\label{fig.phase} }
\end{center}
\end{figure}

Before calculating the RXS spectra, we summarize the magnetic structures
experimentally identified in the Ho metal. The crystal takes an hcp structure.
Below $T_{\textrm{N}}=133$ K, the magnetic moments of the
system order in a helical structure.
The moments on each Ho layer are confined to the basal
plane and are coupled ferromagnetically within each layer.
The interlayer angle of the moment varies
a constant angle per layer.
We call it as "\textit{uniform helical phase}".
A schematic explanation of the orientation of the
moment on each layer of this phase
is given in Fig. \ref{fig.phase} (a).
The number of Ho layers $N$ within one helical pitch is connected to
the modulation vector $\textbf{q}=(0,0,2/N)$ in units of $2\pi/c$.
The $N$ increases on cooling because the helical structure is distorted,
or more precisely, the magnitude of $\textbf{q}$ reduces 
from $\textbf{q}=(0,0,0.271)$ corresponding
to $N \simeq 7$ just below $T_{\textrm{N}}$ 
to $\textbf{q}=(0,0,\frac{1}{6})$ corresponding 
to $N=12$ at $T_{\textrm{C}}=20$ K.
Below $T_{textrm{C}}$, $\textbf{q}$ is locked in 
to $(0,0,\frac{1}{6})$ corresponding to $N=12$
with the emergence of a ferromagnetic component
along the $c$-axis, called as "\textit{conical phase}".
A complication is that the orientation of the
moment within a basal plane does not alters
constantly.\cite{Felcher76,Pechan88}

The magnetic easy axes seem to direct to the centers of six triangles
formed by neighboring atoms.  
The actual structure in the conical phase is that
the twelve Ho layers form six pairs 
of two layers with the moments directing nearly to one of
the centers of six triangles formed by nearest neighbor atoms
as shown in Fig.~\ref{fig.phase} (b).
The bunching angle between the pairs across the
same easy axis is defined as $\gamma$ and
evaluated about $5.8^{\circ}$ in the
$T \rightarrow 0$ limit.

The transition between the uniform helical
and the conical phases is not a
straightforward depending on the conditions such as
the strain present in the sample, the presence
of the chemical impurities, and so on.\cite{Helgesen94}
When the temperature
is above $T_{\textrm{C}}$ but well below
$T_{\textrm{N}}$, the $c$ component of the moment
disappears but the bunching remains.
At the same time, some of the pairs lose their
partner and the remaining moment directs to one of the centers
of six triangles formed by nearest neighbor atoms.
Gibbs \textit{et al}. called this phase
as the "\textit{spin slip phase}".
For example, when one pair experiences a spin slip,
the unit cell of the
system becomes five bunched pairs plus a single
component with $N=11$ (See Fig. \ref{fig.phase} (c)).
When the temperature raises, the value of $\gamma$
increases resulting in the uniform helical phase.
In this paper, we show that the spectral shapes
of the RXS spectra changes drastically when 
the magnetic structure changes from the uniform
helical one to the others.

\section{\label{sect.3}An extension of the RXS formula of 
Hannon \lowercase{\textit{et al}.}}

A resonant process is described as following:
an incident photon with the energy $\hbar \omega$,
wave vector $\textbf{k}$, and polarization
$\mbox{\boldmath{$\epsilon$}}$ is scattered
into the state with the energy $\hbar \omega$,
wave vector $\textbf{k}'$, and polarization
$\mbox{\boldmath{$\epsilon$}}'$ through
the process that the core electron is excited
into an unoccupied level leaving a core hole 
and then recombines with the core hole by emitting 
the photon. 
In a localized electron picture, the scattering amplitude
is well approximated by a sum of the contributions from each site.
Therefore, the RXS amplitude $f_{\textrm{E1}}
(\mbox{\boldmath{$\epsilon$}},
\mbox{\boldmath{$\epsilon$}}',\textbf{k},\textbf{k}';\omega)$
in the $E$1 transition is given by
\begin{equation}
f_{\textrm{E1}}
=\frac{1}{\sqrt{N_0}} \sum_j 
\textrm{e}^{-i \textbf{G}\cdot \textbf{r}_j}
M_j(\mbox{\boldmath{$\epsilon$}},\mbox{\boldmath{$\epsilon$}}';
\omega), \label{eq.amplitude.0} 
\end{equation}
with
\begin{equation}
M_j(\mbox{\boldmath{$\epsilon$}},\mbox{\boldmath{$\epsilon$}}';
\omega) = \sum_{\alpha' \alpha} 
\epsilon_{\alpha}' \epsilon_{\alpha '}
\sum_{\Lambda}
\frac{\langle 0_j | x_{\alpha,j}| \Lambda \rangle
       \langle \Lambda | x_{\alpha',j}|0_j \rangle}
       {\hbar \omega -(E_{\Lambda}-E_0) + i \Gamma},
\label{eq.general}
\end{equation}
within the second order of the photon-electron
interaction in the perturbation theory.
Here the scattering vector is defined 
as $\textbf{G}=\textbf{k}'-\textbf{k}$, and
the number of sites is represented as $N_0$.
The $M_j(\mbox{\boldmath{$\epsilon$}},
\mbox{\boldmath{$\epsilon$}}';\omega)$
is the scattering amplitude of single site $j$ with the 
position vector $\textbf{r}_j$,
where the initial state is represented as $|0_j \rangle$ with the
energy $E_0$ and the intermediate state is given by
$|\Lambda \rangle$ with the energy $E_{\Lambda}$.
The lifetime broadening width of the core hole is
denoted as $\Gamma$. The dipole operators $x_{\alpha}(j)$'s
are defined as $x_1(j)=x$, $x_2(j)=y$, and $x_3(j)=z$
in the coordinate system fixed to the crystal axes
with the origin located at the site $j$.
The RXS intensity 
is proportional to the square of the absolute value of the
scattering amplitude.

The evaluation of eq.~(\ref{eq.general}) is not an easy task, 
because the intermediate states are to be summed up.
We consider the situation that the initial state is described within 
a multiplet of certain total angular momentum $J$, and the degeneracy is 
lifted by the crystal electric field (CEF) and the inter-atomic interaction.
This situation is familiar in many f electron systems.
On the other hand, we neglect the CEF and the inter-atomic interaction 
on the intermediate states. This may be justified 
because the energy dependence of the spectra is mainly determined 
by the multiplets which energy is mulch larger than the CEF and the 
inter-atomic interaction. With assumption, the intermediate states possess 
the rotational symmetry, and thereby the summation over these states are 
rather simply carried out. We could express the scattering amplitude 
at site $j$ in a simple form as
\begin{eqnarray}
 M_j
  &=& \alpha_{\textrm{E1}}^{(0)}(\omega) 
(\mbox{\boldmath{$\epsilon$}}'\cdot \mbox{\boldmath{$\epsilon$}})
- i \alpha_{\textrm{E1}}^{(1)}(\omega) 
(\mbox{\boldmath{$\epsilon$}}'\times \mbox{\boldmath{$\epsilon$}})
\cdot \langle 0_j | \hat{\textbf{J}} | 0_j \rangle 
\nonumber \\
&+& \alpha_{\textrm{E1}}^{(2)}(\omega)
\sum_{\nu=1}^5 
P_{\nu}^{(2)}
(\mbox{\boldmath{$\epsilon$}},\mbox{\boldmath{$\epsilon$}}') 
\langle 0_j | \hat{z}_{\nu} | 0_j \rangle .
\label{eq.amplitude} 
\end{eqnarray}
Here $\alpha_{\textrm{E1}}^{(n)}(\omega)$ represents
the energy profile of rank $n$ operator, which depends only on the initial
state through the magnitude of the angular momentum $J$.
The geometrical factors 
associated with the rank two quantities are
represented as
$P_{\nu}^{(2)}(\mbox{\boldmath{$\epsilon$}},
\mbox{\boldmath{$\epsilon$}}')$.
The components of the (rank two) quadrupole operator
are defined as
\begin{equation}
\hat{z}_{\nu} = \left\{ \begin{array}{lc}
\hat{O}_{x^2-y^2}=
\frac{\sqrt{3}}{2}(\hat{J}_x^2-\hat{J}_y^2), & \nu=1 \\
\hat{O}_{3z^2-r^2}=
\frac{1}{2}[3\hat{J}_z^2-J(J+1)], & \nu=2 \\
\hat{O}_{yz}=
\frac{\sqrt{3}}{2}(\hat{J}_y \hat{J}_z
                  +\hat{J}_z \hat{J}_y), & \nu=3 \\
\hat{O}_{zx}=
\frac{\sqrt{3}}{2}(\hat{J}_z \hat{J}_x
                  +\hat{J}_x \hat{J}_z), & \nu=4 \\
\hat{O}_{xy}=
\frac{\sqrt{3}}{2}(\hat{J}_x \hat{J}_y
                  +\hat{J}_y \hat{J}_x), & \nu=5 \\
\end{array} \right. .
\end{equation}
The detailed derivation of the above expression
including the definitions of quantities
$\alpha_{\textrm{E1}}^{(n)}(\omega)$ and 
$P_{\nu}^{(2)}(\mbox{\boldmath{$\epsilon$}},
\mbox{\boldmath{$\epsilon$}}')$ are
found in our previous papers,\cite{Nagao05,Nagao05.2,Nagao06}
in which 
applications to several $f$ electron systems
and an extension to the $E$2 process are explored too.

The formulae similar to eq. (\ref{eq.amplitude})
are found in the previous 
literatures.\cite{Hannon88,Carra94,Hill96,Lovesey96}
One of the most notable formulae is the one 
derived by Hannon \textit{et al}.,\cite{Hannon88}
which is expressed for the magnetic ordered system as
\begin{equation}
 M_j = (\mbox{\boldmath{$\epsilon$}}' \cdot
\mbox{\boldmath{$\epsilon$}}) f_{0} '(\omega)
-i 
(\mbox{\boldmath{$\epsilon$}}' \times
\mbox{\boldmath{$\epsilon$}}) \cdot \textbf{m}_j 
f_{1} '(\omega)
+(\mbox{\boldmath{$\epsilon$}}' \cdot \textbf{m}_j)
(\mbox{\boldmath{$\epsilon$}} \cdot \textbf{m}_j) 
f_{2} '(\omega), \label{eq.Hannon}
\end{equation}
where $\textbf{m}_j$ is the unit vector in the direction
of the magnetic moment at site $j$.
The energy profiles are represented 
by $f_{n} '(\omega)$'s
describing the contributions from the order $n$ of the
magnetic moment.
Although both are similar in a symmetrical point of view,
our expression is more accurate and superior than that of Hannon \textit{et al}.
in the following reason.

First, higher rank multipole is naturally treated
by our expression, while that of Hannon \textit{et al}.
was expressed in terms of the expectation values of the dipole moment operator.
To make the comparison easy, we rewrite the last term of 
eq.~(\ref{eq.amplitude}) as
\begin{eqnarray}
 \alpha_{\textrm{E1}}^{(2)}(\omega) & \sum_{\nu=1}^5 &
P_{\nu}^{(2)}
(\mbox{\boldmath{$\epsilon$}},\mbox{\boldmath{$\epsilon$}}') 
\langle 0_j | \hat{z}_{\nu} | 0_j \rangle \nonumber \\
 &=& \alpha_{\textrm{E1}}^{(2)}(\omega) \frac{3}{2}
 \left[\frac{1}{2}\langle(\mbox{\boldmath{$\epsilon$}}\cdot{\bf J})
 (\mbox{\boldmath{$\epsilon$}}'\cdot{\bf J})
+ (\mbox{\boldmath{$\epsilon$}}'\cdot{\bf J})
 (\mbox{\boldmath{$\epsilon$}}\cdot{\bf J})\rangle
 -\frac{J(J+1)}{3}(\mbox{\boldmath{$\epsilon$}}'\cdot
 \mbox{\boldmath{$\epsilon$}})\right].
\end{eqnarray}
In comparison with Hannon \textit{et al}.'s,
the expectation value of the product of the
operators is generally different from the
product of the expectation values of the operators.
Our expression is applicable to other multipole ordered phases
without the local magnetic moment.
The numerical amount of the difference 
between eqs.~(\ref{eq.amplitude}) and (\ref{eq.Hannon})
depends on the system considered in the magnetic ordered phase.
In the present case of holmium, the difference is
relevant to the analysis of the second harmonics satellite spot, 
which brings about a quantitative outcome when we use eqs. 
(\ref{eq.amplitude}) and (\ref{eq.Hannon}) as a fitting function.

Second, our expression gives the spectral shape of 
the RXS intensity as a function of the incident photon energy,
without relying on the so-called \textit{fast collision} (FC) approximation.
Recent development of the high resolution
in energy enables the detection of the reliable
spectral profiles as a function of the incident
photon energy, which increases the value of our treatment
including the energy dependence. 

\section{\label{sect.4}Procedure to evaluate
the RXS amplitude in H\lowercase{o}}

Before going to the detailed analysis of RXS spectra in Ho,
we briefly explain how to evaluate eq. (\ref{eq.amplitude})
based on a localized electron picture.

\subsection{\label{sect.4.1}Evaluation of $\langle 0_j|{\bf J}|0_j\rangle$
and $\langle 0_j|z_{\nu}|0_j\rangle$}

The holmium in the solid behaves, to a good approximation,
as trivalent ion with the $(4f)^{10}$
configuration equivalent to $\underline{f}^4$ in the hole picture. 
We expect  from Hund's rule the states of $^{5}\textrm{I}_8$ 
as the ground multiplet, which is denoted as $|J=8, J_z\rangle$.
The degeneracy of this multiplet would be lifted by the intersite interaction,
resulting in the magnetically ordered states.
Therefore, the magnetic wavefunctions at each site are expanded 
in terms of $|J=8,J_z\rangle$'s. It is known that
the local magnetic moment is as large as about $10 \mu_{\textrm{B}}$
in the conical phase. This value is equivalent to the maximum one expected 
from the saturation value of $^{5}\textrm{I}_8$. 
It is also known that the local magnetic moment is about 70 \% of 
the saturated value around $T/T_{\textrm{N}} \sim 0.6$ in the uniform 
helical phase. Accordingly, we assume that the magnetic moment at each Ho site 
is saturated, since the spectral shape of the energy profile
$\alpha_{\textrm{E1}}^{(n)}(\omega)$ is not sensitive to the value.
Therefore, the wavefunction is given by $|J=8,J_z=8\rangle$ at each site
in the local coordinate frame where the $z$-axis is pointing
to the direction of the local magnetic moment.

Let $(\theta_n,\beta,0)$ be
the Euler angle connecting the local coordinate frame to
the crystal fixed frame for the $n$-th Ho layer,
and $|\theta_n\rangle$ be the corresponding magnetic state.
Then, the expectation values of the dipole and the quadrupole operators 
defined by the crystal fixed frame are evaluated as
\begin{equation}
\langle \theta_n | \left\{ \begin{array}{c}
\hat{J}_x \\
\hat{J}_y \\
\hat{J}_z \\
\end{array} \right\} | \theta_n \rangle
= J
\left\{ \begin{array}{c}
\sin \beta \cos \theta_n \\
\sin \beta \sin \theta_n \\
\cos \beta \\
\end{array}
\right\}, \label{eq.eigen1}
\end{equation}
and
\begin{equation}
\langle \theta_n |
\left\{ \begin{array}{c}
\hat{O}_2^2 \\
\hat{O}_2^0 \\
\hat{O}_{yz} \\
\hat{O}_{zx} \\
\hat{O}_{xy} \\
\end{array} \right\} | \theta_n \rangle
= \frac{\sqrt{3}}{4} J(2J-1)
\left\{ \begin{array}{c}
\sin^2 \beta \cos (2\theta_n) \\
\frac{1}{\sqrt{3}} ( 3 \cos^2 \beta -1 ) \\
\sin (2\beta) \sin \theta_n \\
\sin (2\beta) \cos \theta_n \\
\sin^2 \beta \sin (2\theta_n) \\
\end{array}
\right\}. \label{eq.eigen2}
\end{equation}
These expressions are used for
$\langle 0_j|{\bf J}|0_j\rangle$ and $\langle 0_j|z_{\nu}|0_j\rangle$
in eq.~(\ref{eq.amplitude}).

\subsection{\label{sect.4.2}Evaluation of $\alpha^{(i)}_{E1}(\omega)$}

Energy profiles $\alpha^{(i)}_{E1}(\omega)$ ($i=0,1,2$) depend on the initial 
state only through $J$ of the ground multiplet but little on the details
of the magnetic orders. They also depend on the intermediate states.
For the ground multiplet, we consider the intra-atomic Coulomb and the spin 
orbit (SO) interactions, where the parameters needed to evaluate the above 
interactions such as the Slater integrals for the Coulomb interaction
and the SO coupling constant are calculated within 
the Hartree-Fock (HF) approximation.\cite{Cowan81,ComScreen}
Representing the Hamiltonian with the bases spanned by the $\underline{f}^4$
configuration, we diagonalize the Hamiltonian matrix.
We obtain the degenerated lowest seventeen states
corresponding to $J=8$.
In the intermediate states, the electronic configuration becomes 
$(3d)^9(4f)^{11}$ equivalent to $\underline{d}^1 \underline{f}^3$.
We repeat the similar procedure as have done to prepare the ground multiplet.
The Hamiltonian describing the intermediate states
takes full account of
the intra-atomic Coulomb and the SO interactions
among the $(3d)^9(4f)^{11}$ configuration.
By diagonalizing numerically the Hamiltonian matrix represented with the bases
spanned by the $\underline{d}^1 \underline{f}^3$ configuration.
We obtain the set of the intermediate states.
These states are sufficient to evaluate the energy profiles 
$\alpha^{(i)}_{E1}(\omega)$.

\section{Analysis of RXS for H\lowercase{o}\label{sect.5}}
Since all the Ho atoms within the same layer give a
same amount of contribution, the summation over
site $j$ is replaced by that over $n$.
The amplitude for the $m$-th satellite spot is given by
\begin{eqnarray}
f_{\textrm{E1}} &\propto& 
(\mbox{\boldmath{$\epsilon$}}'\cdot \mbox{\boldmath{$\epsilon$}})
\alpha_{\textrm{E1}}^{(0)}(\omega)
\sum_{n=1}^{N} 
\textrm{e}^{-i \textbf{G}_m \cdot \textbf{r}_n}
\nonumber \\
&-& 
i \alpha_{\textrm{E1}}^{(1)}(\omega)
(\mbox{\boldmath{$\epsilon$}}'\times \mbox{\boldmath{$\epsilon$}})
\cdot \left[
\sum_{n=1}^{N} 
\textrm{e}^{-i \textbf{G}_m \cdot \textbf{r}_n}
\langle \theta_n |\hat{\textbf{J}} | \theta_n \rangle 
\right] \nonumber \\
&+&
\alpha_{\textrm{E1}}^{(2)}(\omega) 
\sum_{\nu=1}^{5} P_{\nu}^{(2)}
(\mbox{\boldmath{$\epsilon$}}', \mbox{\boldmath{$\epsilon$}})
\left[ \sum_{n=1}^{N} 
\textrm{e}^{-i \textbf{G}_m \cdot \textbf{r}_n}
\langle \theta_n | \hat{z}_{\nu} | \theta_n \rangle \right],
\nonumber \\
\label{eq.amplitude.3}
\end{eqnarray}
where the wavevector of the $m$-th satellite spot is
defined by
\begin{equation}
\textbf{G}_m = \left( 0, 0,\tau_m \right),
\quad \tau_m=m \frac{2}{N}
\label{eq.Gdef}
\end{equation}
which is measured in units of
$\frac{2\pi}{c}$.
The position vector of the $n$-th Ho layer is
represented as $\textbf{r}_n=(0,0,\frac{c}{2} n)$.

Now we describe $\theta_n$ in various magnetic ordered phase.
In the conical phase, it is given by
\begin{equation}
\theta_n = \left\{ \begin{array}{ll}
2\pi\frac{n}{N} + \phi_0 - \gamma &
 \textrm{for} \ n= \ \textrm{odd} \\
2\pi\frac{n-1}{N} + \phi_0 + \gamma &
 \textrm{for} \ n= \ \textrm{even} \\
\end{array} \right. , \label{eq.angle}
\end{equation}
where $\phi_0$ is a constant and taken to be zero.
The number of the holmium layers within a
helical pitch is $N=12$.
In the uniform helical phase, we could put
$N$ as an arbitrary integer, $\phi_0$ as an arbitrary angle, 
and $\gamma=\frac{\pi}{N}$ in eq. (\ref{eq.angle}).
Thereby we have $\theta_n$ for an arbitrary integer $n$,
\begin{equation}
\theta_n = \pi \frac{2n-1}{N} + \phi_0.
\label{eq.angle.1}
\end{equation}
In the spin slip phase, we take a case of $N=11$ system as an example,
since the distribution of $\theta_n$'s show a wide diversity.
The unit cell consists of the five pairs of bunched doublet
and a single layer as schematically
shown in Fig. \ref{fig.phase} (c).
Within the one helical pitch (for
$n=1, 2, \cdots, N$),  $\theta_n$'s are
expressed as
\begin{equation}
\theta_n = \left\{ \begin{array}{ll}
2\pi\frac{n}{N+1} + \phi_0 - \gamma &
 \textrm{for} \ n=1, 3, \cdots, N-2 \\
2\pi\frac{n-1}{N+1} + \phi_0 + \gamma &
 \textrm{for} \ n=2,4, \cdots, N-1 \\
\frac{2\pi}{N+1} & \textrm{for} \ n=N \\
\end{array} \right. , \label{eq.angle.2}
\end{equation}
with $\phi_0=\frac{\pi}{2}$. 
Note that eq. (\ref{eq.angle.2}) is valid only
for $n=1, \cdots, N$, which is enough for the present
analysis.

With the use of eqs. (\ref{eq.eigen1}) and (\ref{eq.eigen2})
together with eqs.~(\ref{eq.angle}), (\ref{eq.angle.1}), and 
(\ref{eq.angle.2}), the calculation of eq. (\ref{eq.amplitude.3})
results in evaluating the summations 
of the types
\begin{eqnarray}
& &
 \sum_{n=1}^{N} \textrm{e}^{-i\textbf{G}_m \cdot \textbf{r}_n}
\times
\left\{ \begin{array}{c}
\cos ( m' \theta_n)\\
\sin ( m' \theta_n)\\
\end{array} \right\} \nonumber \\
&=&
\frac{N}{2} \textrm{e}^{i m\left( \phi_0-\frac{\pi}{N} 
\right)} 
\sum_{\ell}
\cos \left[ \left(m -\frac{\ell N}{2} \right) \gamma 
-\frac{m\pi}{N} \right] 
\nonumber \\
&\times& \textrm{e}^{i \frac{\ell N}{2}\left( 
\frac{2 \pi}{N} -\phi_0 \right)}
\left[ \delta_{m+m',\frac{\ell N}{2}}
\left\{ \begin{array}{c}
1 \\
i \\
\end{array} \right\}
+\delta_{m-m',\frac{\ell N}{2}}
\left\{ \begin{array}{c}
1 \\
(-i) \\
\end{array} \right\} \right], \nonumber \\
\label{eq.sum}
\end{eqnarray}
where $\ell$ is an arbitrary integer.
Note that eq. (\ref{eq.sum}) has no contribution when
$\ell$ is a odd integer in the uniform helical
phase, since the cosine factor becomes 
$\cos\left(\frac{\ell\pi}{2}\right)$ with 
$\gamma=\frac{\pi}{N}$.
In the case of the conical phase, eq. (\ref{eq.sum}) is
applicable for $m' \geq 1$ and
when $m'=0$, the summation is trivial as
$\sum_{n=1}^{N} \textrm{e}^{-i \textbf{G}_m \cdot \textbf{r}_n}
=N \delta_{m,0}$.

\subsection{\label{sect.5.1}Uniform helical phase}
In the uniform helical phase, the observed value
of $N$ is larger than about seven.
Then, only the term proportional to 
$\delta_{m-m',\frac{\ell N}{2}}$ with $\ell=0$ in eq.
(\ref{eq.sum}) is relevant
for the description of the $E$1 process in this phase.
Practically, eq. (\ref{eq.sum}) is rewritten as
\begin{eqnarray}
& &
 \sum_{n=1}^{N} \textrm{e}^{-i\textbf{G}_m \cdot \textbf{r}_n}
\times
\left\{ \begin{array}{c}
\cos ( m' \theta_n)\\
\sin ( m' \theta_n)\\
\end{array} \right\} \nonumber \\
&\rightarrow&
\delta_{m,m'} \frac{N}{2}
\textrm{e}^{i m\left( \phi_0-\frac{\pi}{N} \right)}
\cos \left[ m \left(\gamma-\frac{\pi}{N} \right) 
 \right] 
\left\{ \begin{array}{c}
1 \\
(-i) \\
\end{array} \right\}. \nonumber \\
\label{eq.sum.1}
\end{eqnarray}
By substituting eq. (\ref{eq.sum.1}) into 
eq. (\ref{eq.amplitude.3}), 
and fixing $\beta=\gamma=\frac{\pi}{2}$,
we obtain the RXS amplitude at the first satellite spot as
\begin{equation}
f_{\textrm{E1}} \propto -i
\frac{N}{2} \textrm{e}^{i\left( \phi_0-\frac{\pi}{N} \right)}
 J 
[(\mbox{\boldmath{$\epsilon$}}' \times
\mbox{\boldmath{$\epsilon$}})_x
-i (\mbox{\boldmath{$\epsilon$}}' \times
\mbox{\boldmath{$\epsilon$}})_y ]
 \alpha_{\textrm{E1}}^{(1)}(\omega).
\label{eq.hel.1}
\end{equation}
Note that the contributions from the quadrupole operator
disappear in the uniform helical phase 
at the first satellite spot.
Similarly, the amplitude at the second satellite spot
is given by,
\begin{equation}
f_{\textrm{E1}} \propto
\frac{N}{2} \frac{\sqrt{3}J(2J-1)}{4}
\textrm{e}^{2i\left( \phi_0-\frac{\pi}{N} \right)} 
[ P_{1}^{(2)}-i P_{5}^{(2)}]
 \alpha_{\textrm{E1}}^{(2)}(\omega),
\label{eq.hel.2}
\end{equation}
where arguments are omitted  in the geometrical factor
$P_{\lambda}^{(\nu)}$.
As already known,\cite{Gibbs88,Hannon88,Hill96}
the spectrum at the $m$-th satellite spot
consists of pure rank $m$ profile.
The final forms of the scattering intensities are
derived by substituting the geometrical factors
into eqs. (\ref{eq.hel.1}) and (\ref{eq.hel.2}).
We relegate both the detail of the derivation
and the results to Appendix \ref{app.A}.
Here, we mention an important feature the results possess
that the intensities exhibit no $\psi$ dependence
at both satellite spots with the choice of the scattering
vector as the form in eq. (\ref{eq.Gdef}).

\subsection{\label{sect.5.2}Conical phase}
In the conical phase, $N$ and $\phi_0$ are
fixed to be twelve and zero, respectively.
A notable difference compared with the uniform helical
phase is that 
the terms with $\ell = \pm 1$ becomes to be relevant in
eq. (\ref{eq.sum}) in the conical phase. 
Let us state the result with $\ell=0$ and $\pm 1$, in turn.

First, the relevant terms in eq. (\ref{eq.sum})
with $\ell=0$ are summarized as follows.
At the first satellite spot ($m=1$), the amplitude
is written as
\begin{eqnarray}
f_{\textrm{E1}} &\propto&
6 \textrm{e}^{-i\frac{\pi}{12}}
\cos \left(\gamma-\frac{\pi}{12} \right) J \sin \beta \nonumber \\
&\times& \left[
-i \{(\mbox{\boldmath{$\epsilon$}}' \times
\mbox{\boldmath{$\epsilon$}})_x
-i (\mbox{\boldmath{$\epsilon$}}' \times
\mbox{\boldmath{$\epsilon$}})_y
\}
 \alpha_{\textrm{E1}}^{(1)}(\omega) \right. \nonumber \\
&+ & \left. \frac{\sqrt{3}(2J-1)\cos\beta}{2}
\{P_{4}^{(2)} -i P_{3}^{(2)} \} 
\alpha_{\textrm{E1}}^{(2)}(\omega)\right].
\label{eq.con.1}
\end{eqnarray}
The amplitude is a mixture of 
$\alpha_{\textrm{E1}}^{(1)}(\omega)$ and 
$\alpha_{\textrm{E1}}^{(2)}(\omega)$,
while the one is pure $\alpha_{\textrm{E1}}^{(1)}(\omega)$
in the uniform helical phase.
This indicates that there is a possibility
to observe the variation of the spectral shape
of the first satellite spot when the underlying magnetic
structure changes.
However, such a possibility is practically denied since
$|\alpha_{\textrm{E1}}^{(1)}(\omega)|$
is much larger than 
$|\alpha_{\textrm{E1}}^{(2)}(\omega)|$
as is confirmed numerically in the next section.
 
The amplitude at the second satellite spot ($m=2$)
is given by,
\begin{eqnarray}
f_{\textrm{E1}} &\propto&
6 \textrm{e}^{-i\frac{\pi}{6}}
\cos \left(2\gamma-\frac{\pi}{6} \right) \sin^2 \beta \nonumber \\
&\times&  \frac{\sqrt{3}}{4}J(2J-1)
[ P_{1}^{(2)}-i P_{5}^{(2)}]
 \alpha_{\textrm{E1}}^{(2)}(\omega).
\label{eq.con.2}
\end{eqnarray}
Note that both amplitudes do not include a contribution
from $\langle \theta_n|\hat{J}_z|\theta_n \rangle$
despite the fact that the conical phase is characterized
by the non-zero value of it.

Next, we proceed to summarize the result expected
from the terms with $\ell=\pm 1$ in eq. (\ref{eq.sum}).
This raises a possibility that signals from the
higher-order satellites may be detected. 
The possibility is linked to the distortion of the
direction of the magnetic moment within the basal plane.
Koehler \textit{et al}. have reported that
such higher-order magnetic satellites as fifth- and 
seventh-order spots were present in the 
neutron scattering.\cite{Koehler66}
In the $E$1 process of R(M)XS, we see the possibility
that the intensities from the 
fourth-, eighth-order in addition to fifth- and seventh-order
satellites are detectable. 
The possibility of detecting the former two satellite
enable RXS to be a complement to neutron scattering. 

The fifth and seventh satellite spots have contributions
from rank one and two profiles.  The intensities are
proportional to
\begin{eqnarray}
 &\propto&
\cos^2 \left( \gamma+\frac{5\pi}{12} \right) 
\left|
-i \{(\mbox{\boldmath{$\epsilon$}}' \times
\mbox{\boldmath{$\epsilon$}})_x
\pm i (\mbox{\boldmath{$\epsilon$}}' \times
\mbox{\boldmath{$\epsilon$}})_y
\}
 \alpha_{\textrm{E1}}^{(1)}(\omega) \right. \nonumber \\
&+ & \left. \frac{\sqrt{3}(2J-1)\cos\beta}{2}
\{P_{4}^{(2)} \pm i P_{3}^{(2)} \} 
\alpha_{\textrm{E1}}^{(2)}(\omega)\right|^2,
\end{eqnarray}
where $m= \pm 5$ and $\pm 7$. The upper and lower signs in the
equation are taken in accordance with that of $m$.
The intensities at the fourth and eighth
satellite spots have contribution from rank two 
profile, being proportional to
\begin{equation}
\propto
\cos^2 \left( 2\gamma + \frac{\pi}{3} \right)
| P_{1}^{(2)} \pm i P_{5}^{(2)}|^2
| \alpha_{\textrm{E1}}^{(2)}(\omega)|^2,
\end{equation}
where $m =\pm 4$ and $\pm 8$.
The upper and lower signs in the
equation are taken in accordance with that of $m$.
The detectabilities of these intensities at the 
higher order satellite spots are determined by the 
cosine factors and the geometrical factors 
appeared in the above expressions.
If we assume $\gamma=5.8^{\circ}$, the former factor
gives 0.08, 0.19, 0.48, and 0.64
with $m=\pm4, \pm 5, \pm 7$, and $\pm 8$, respectively.
Near Ho $M_5$ edge, the Bragg angle
$\theta^{(m)}=33.1^{\circ}, 43.1^{\circ}$, and $73.0^{\circ}$
with respect to $m=\pm 4, \pm 5$, and $\pm 7$, respectively.
With the substitution of these parameter values,
we can conclude that the intensity from the
$\alpha_{\textrm{E1}}^{(2)}(\omega)$ contribution
at the Ho $M_5$ fourth satellite spot 
in the $\pi-\sigma'$ (or $\sigma-\pi'$) channel is 
large enough to
be detected. Actually, the intensity is expected
as nearly the same as that observed at the second satellite
spot by Ott \textit{et al}.\cite{Ott06}
Finally, we end this subsection with a comment that
the intensities present no azimuthal angle 
dependence with the choice of 
$\textbf{G}_m$ as eq. (\ref{eq.Gdef}) in the conical phase 
as shown in App. \ref{app.A}.

\subsection{\label{sect.5.3}Spin slip phase}
Since the magnetic structure
of this phase is filled with variation, we restrict ourself
on the case of $N=11$ system depicted 
in Fig. \ref{fig.phase} (c).
The distribution of $\theta_n$'s is irregular
at the spin-slipped layer in this phase. 
Then, the summation over $n$ similar to the left hand
side of eq. (\ref{eq.sum}) is not results in the right hand
side of it.
As a consequence, there remains the terms for $m \neq m'$.
Two prominent properties are readily anticipated.
First, the intensities at the first and second satellite
spots exhibit $\psi$ dependence.
Second, the profiles consist of the mixture
of $\alpha_{\textrm{E1}}^{(1)}(\omega)$
and $\alpha_{\textrm{E1}}^{(2)}(\omega)$ at
both spots, which leads to the situation that
both spectra are dominated by 
$\alpha_{\textrm{E1}}^{(1)}(\omega)$.
The latter point at the first satellite spot is
equivalent to the situation observed in the
conical phase, eq. (\ref{eq.con.1}).
It brings about no practical change of the spectral
shape since $\alpha_{\textrm{E1}}^{(1)}(\omega)$ always
dominates the entire spectrum.
On the other hand, the RXS amplitude at the second satellite
spot is completely new one, since those in the
conical and uniform helical phases are made
of pure $\alpha_{\textrm{E1}}^{(2)}(\omega)$.

\subsection{\label{sect.5.4}Clockwise or anti-clockwise}
In the helical (or conical) magnetic structure of
holmium sample, usually, there exist two domains
distinguished by the orientation of the winding
of the magnetic moment as clockwise and anti-clockwise. 
The latter is defined
such that the angle $\theta_n$ of the moment 
increases anti-clockwise
when the number of Ho layer $n$ increases as shown
in Fig. \ref{fig.phase}, while the former the opposite.
Since experimentally observed RXS intensities are expected to 
be the domain averaged quantities, we 
should mention what is anticipated from 
the clockwise and anti-clockwise domains. 

By checking eqs. (\ref{eq.hel.1}) and (\ref{eq.hel.2}),
it turns out that
the RXS intensity shows no domain dependence in the
uniform helical phase. 
The same is true at the second satellite spot in the
conical phase.  The intensity at the first satellite spot,
however, shows domain dependence when the scattering
vector is not the $(0,0,\tau_1)$-type. 
For example, the intensities with 
$\textbf{G}=(2,1,\tau_1)$ may give domain dependence.
Fortunately, such $\textbf{G}$ is impossible
to achieve at the Ho $M_5$ absorption edge.
When the experiment is possible for this scattering
vector at the $L_3$ edge, the domain consideration
will be necessary interpreting the result.
As a result, it is no need to worry about the domain consideration when analyzing the conical and the uniform
helical phases.
As for the spin slip phase, the spectrum shows the
domain dependence, at least in a qualitative sense,
regardless of the scattering vector.
So we take domain average when numerical evaluation is
carried out in the spin slip phase in the next section.

\section{\label{sect.6}Numerical results}
We are now in a position to calculate the
RXS spectra in order to compare with the experimental 
result.   Before the comparing process goes forward, we
first expect that the sample used in the measurement
is in the uniform helical phase, not in the 
conical nor spin slip phases, because Ott \textit{et al}.
explicitly did not reported such information nor temperature
the measurement were performed.\cite{Ott06,Com2}
The data were obtained from the Ho-metal thin film
with 16 mono-layers (ML).
The critical temperature $T_{\textrm{N}}$ of the
16 ML sample is about eighty percent of the bulk
one\cite{Weschke04}
and the first harmonics satellite peak
is observed around $\tau_1 \simeq 0.3$ \AA$^{-1}$.\cite{Ott06}
This value of $\tau_1$ suggests that the temperature
is $T=0.55 T_{\textrm{N}}$ inferred from Fig. 4 of ref. 
\citen{Ott06}.   In bulk Ho, the ratio
$T/T_{\textrm{N}}=0.55$ corresponds to $T \simeq 73$ K
and $\tau_1=0.252$ \AA$^{-1}$ with $N \simeq 9$.
On the other hand, if we apply the value of $\tau_1$ to the
bulk leading to $T \simeq 115$ K with 
$N \simeq 7.5$.\cite{Gibbs85}
Since the uniform helical phase is transformed 
into the spin slip phase below about 50 K with
$\tau_1 \simeq 0.22 \textrm{\AA}$ in the bulk,
both estimations lead us to a conclusion that the
experiment were performed in the uniform
helical phase of the sample as anticipated.
Later, we will lend a further support that
this assumption is sustained
when we calculate the second harmonics satellite
spectrum. We show if the sample was in the spin slip phase,
the spectral shape might be completely different one.
In the following, we assume $N=8$, i.e.,
four unit cells are involved in one helical
pitch.
We emphasize that the choice of $N$ does not alter
the spectral shapes.

\begin{figure}[h]
\begin{center}
\includegraphics[width=8.0cm]{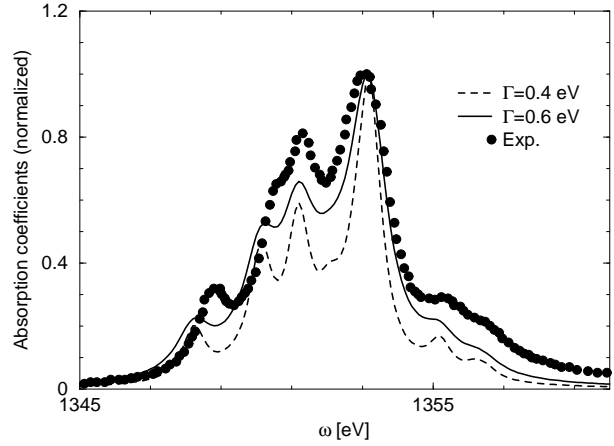}%
\caption{\label{fig.abs.M5}
Absorption coefficient at the Ho $M_{5}$ edge
corresponding to 
$-\textrm{Im}\alpha_{E1}^{(0)}(\omega)$.
The solid and dashed lines are obtained with
$\Gamma=0.6$ and $0.4$ eV, respectively.
Filled circles are the experimental data.\cite{Ott06}}
\end{center}
\end{figure}

\subsection{Absorption coefficients}
We discuss the absorption coefficient $A(\omega)$
in the $E$1 transition, which is described as
\begin{equation}
A(\omega) \propto \Gamma
\sum_{j}\sum_{\Lambda} \sum_{\alpha}
\frac{|\langle \Lambda|x_{j,\alpha}(j)|\psi_0 \rangle|^2}
      {(\hbar \omega - E_{\Lambda})^2 + \Gamma^2}.
\end{equation}
It is obtained from eq. (\ref{eq.amplitude})
by the relation
\begin{equation}
A(\omega) \propto -\textrm{Im} 
\alpha_{\textrm{E1}}^{(0)}(\omega),
\end{equation}
where $\textrm{Im} X$ represents the imaginary part of $X$.
Although the absorption spectrum at the Ho $M_5$ edge
was already investigated by others,\cite{Thole85,Schille93}
here, we utilize it in three-fold purposes;
the determination of the origin of the energy, 
that of the value of $\Gamma$, and the
justification of the atomic picture.

The calculated result is
shown in Fig. \ref{fig.abs.M5}.
The origin of the energy is adjusted such that
the maximum point is located at the
experimental one, i.e., at 1353.2 eV 
and is fixed throughout this paper.
Since the peak structure
of the spectrum shows $\Gamma$ dependence,
several values of $\Gamma$ are examined.
As shown in Fig. \ref{fig.abs.M5},
we find $\Gamma \sim 0.6$ eV gives a better choice.
We do not try, however, to obtain the best fit to the data
considering the simple treatment based on the atomic picture.

This value of $\Gamma$ is consistent
with the previous work for the absorption spectra.\cite{Thole85}
We confirm that 
the small change of the $\Gamma$ value around
0.6 eV has no effect on the peak positions.
Our result reproduces the characteristic multi-peak
structure exhibited by the experimental data.\cite{Ott06}
The total width of the spectrum is governed by the
multiplet splitting of the energy of the
intermediate states.
Aside from the peak position around 1348.5 eV in the
experiment, we conclude that
the atomic treatment gives a reasonable result.
Fortunately, we shall confirm in the next subsection
that this discrepancy causes no practical effect on
the investigation of the RXS spectra,
since the intensities
near 1348.5 eV at the first and second satellite spots
are weak and merely the tail part of the spectra.

\subsection{First and second harmonics satellite}
The RXS spectral intensities at the first and second harmonics
satellite spots are 
proportional to
the dipole ($|\alpha_{\textrm{E1}}^{(1)}|^2$)
and the quadrupole 
($|\alpha_{\textrm{E1}}^{(2)}|^2$) profiles, respectively.
The analytical forms are derived by substituting
$\beta=\frac{\pi}{2}$ and $\gamma=\frac{\pi}{N}$ into
eqs. (\ref{eq.App.A4}) and (\ref{eq.App.A5})
as shown below.
The intensity at the first satellite spot is proportional to
\begin{equation}
I^{(1)}\propto \left(
J \cos \theta^{(1)} \right)^2
|\alpha_{\textrm{E1}}^{(1)}(\omega) |^2
\left\{
\begin{array}{lc}
0 & \textrm{for} \ \sigma-\sigma'\\
1 & \textrm{for} \ \sigma-\pi'\\
1 & \textrm{for} \ \pi-\sigma'\\
4 \sin^2 \theta^{(1)} & \textrm{for} \ \pi-\pi'\\
\end{array}
\right. , \label{eq.helical.G1}
\end{equation}
and the one at the second satellite spot is proportional to
\begin{equation}
I^{(2)} \propto 
\left[\frac{3J(2J-1)}{8} \right]^2
 |\alpha_{\textrm{E1}}^{(2)}(\omega) |^2
\left\{
\begin{array}{lc}
1 & \textrm{for} \ \sigma-\sigma'\\
\sin^2 \theta^{(2)} & \textrm{for} \ \sigma-\pi'\\
\sin^2 \theta^{(2)} & \textrm{for} \ \pi-\sigma'\\
\sin^4 \theta^{(2)} & \textrm{for} \ \pi-\pi'\\
\end{array}
\right. . \label{eq.helical.G2}
\end{equation}

\begin{figure}[h]
\begin{center}
\includegraphics[width=8.0cm]{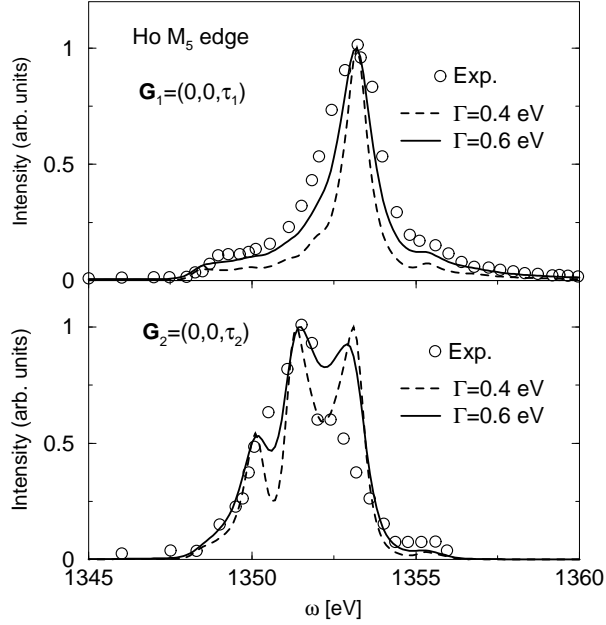}%
\caption{\label{fig.M5}
The RXS spectra near the Ho $M_5$ edge at the first
 (upper) and second (lower) satellite spots.
The curves for $\Gamma=0.4$ eV and $0.6$ eV are
distinguished by the dashed and solid lines, respectively.
Open circles represent 
the experimental result.\cite{Ott06}}
\end{center}
\end{figure}

As seen from the above expressions, the RXS intensities
show no $\psi$ dependence.
Note that spectral shapes have no 
polarization dependence. 
In order to compare our results with the experimental
ones, we consider
the spectra in the $\pi-\sigma'$ and $\pi-\pi'$ channels.
The ratio $I^{(1)}/I^{(2)}$ is common
in both channels
with an aid of the relation $\sin \theta^{(2)}=2
\sin\theta^{(1)}$.
Thus, we can fix $\psi=0$ and restrict in the $\pi-\sigma'$
channel hereafter.
The calculated spectra of $I^{(1)}$ and $I^{(2)}$
are shown in the upper and lower
panels, respectively, of Fig. \ref{fig.M5}
together with the experimental data.\cite{Com1}

As for the $|\alpha_{\textrm{E1}}^{(1)}|^2$
from the first satellite spot with 
$\textbf{G}_1=(0,0,\tau_1)$, the
peak position of the spectrum
appears at the same position
of the absorption coefficient, which is 
coincide with the experiment.
The entire spectral shape with $\Gamma=0.6$ eV
agrees remarkably well with the experimental one
including the kink in the low energy tail and the
hump in the high energy tail.

As for the $|\alpha_{\textrm{E1}}^{(2)}|^2$
from the second satellite spot with 
$\textbf{G}_2=(0,0,\tau_2)$, 
the maximum height intensity is located at the 
energy 1.8 eV lower than that of the absorption as
the same position as the experiment.
The entire spectral width and the shape of the
tail parts show quite similar to the experiment, too. 
On the other hand, however, 
a prominent discrepancy can be found
between the theory and the experiment.
The peak around 1353 eV found in the
the calculated result is absent in the experiment.
The reason of this discrepancy is still unclear.
From the calculated value of the ratio
$I^{(1)}/I^{(2)}$, we deduce a ratio
\begin{equation}
 \frac{|\alpha_{\textrm{E1}}^{(1)}(\omega)|}
{|\alpha_{\textrm{E1}}^{(2)}(\omega)|}
\simeq 27.7, \label{eq.ratio}
\end{equation} 
with $\Gamma=0.6$ eV. This large ratio allows us
to expect the dominance of 
$\alpha_{\textrm{E1}}^{(1)}(\omega)$ in the entire spectral
shape when both profiles are mixed.

\begin{figure}[h]
\begin{center}
\includegraphics[width=8.0cm]{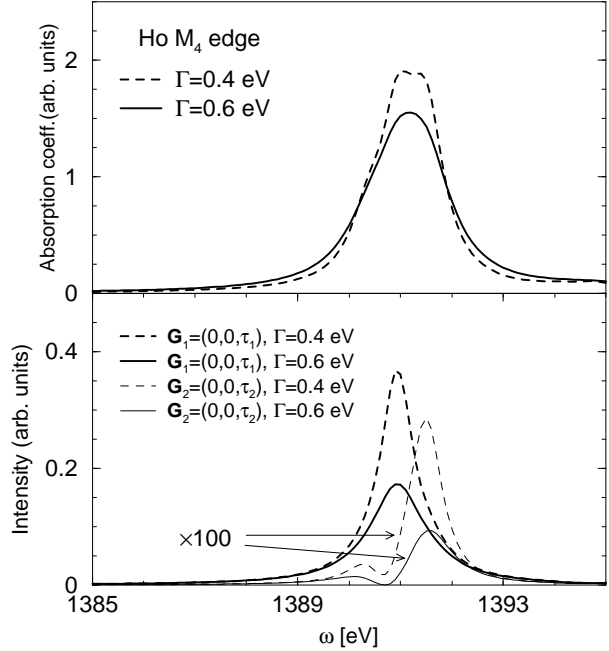}%
\caption{\label{fig.M4}
Absorption coefficient and
the RXS spectra at the Ho $M_4$ edge.
Upper: the absorption coefficients with $\Gamma=0.6$ eV and
$0.4$ eV are shown by the solid and dashed lines, respectively.
Lower: the RXS spectra at the first and second harmonics 
satellite spots are represented by the thick and thin lines,
respectively.
The values of $\Gamma=0.4$ eV and $0.6$ eV are
distinguished by the dashed and solid lines, respectively.
}
\end{center}
\end{figure}

\subsection{$M_4$ edge}
We carry out the similar calculations at the Ho $M_4$
edge as those at the $M_5$ edge.
The absorption coefficient is shown in the upper panel
of Fig. \ref{fig.M4}.
It shows a single peak structure and
the peak height of the $M_4$ spectrum is about 0.175
times the one at the $M_5$ edge.
These observations agree with the experiment.\cite{Thole85}
The RXS spectra expected from the first and second
harmonics satellite spots are shown 
in the lower panel of Fig. \ref{fig.M4}.
The maximum intensity of the second satellite spectrum
is found at the energy about 0.6 eV higher
than that of the first satellite spectrum.
This tendency 
is contrary to the case observed at the $M_5$ edge
where the former is found at the energy about
1.8 eV lower than the latter.
The peak intensities at the 
$M_4$ edge are about two times weaker than
those at the $M_5$ edge.  Since the order of the magnitude
of the spectrum of the first satellite spot at the $M_4$ edge
is roughly the same order 
as that of the second satellite spot at the $M_5$ spot,
the former may be detectable experimentally.

\subsection{spin-slip phase}
\begin{figure}[h]
\begin{center}
\includegraphics[width=8.0cm]{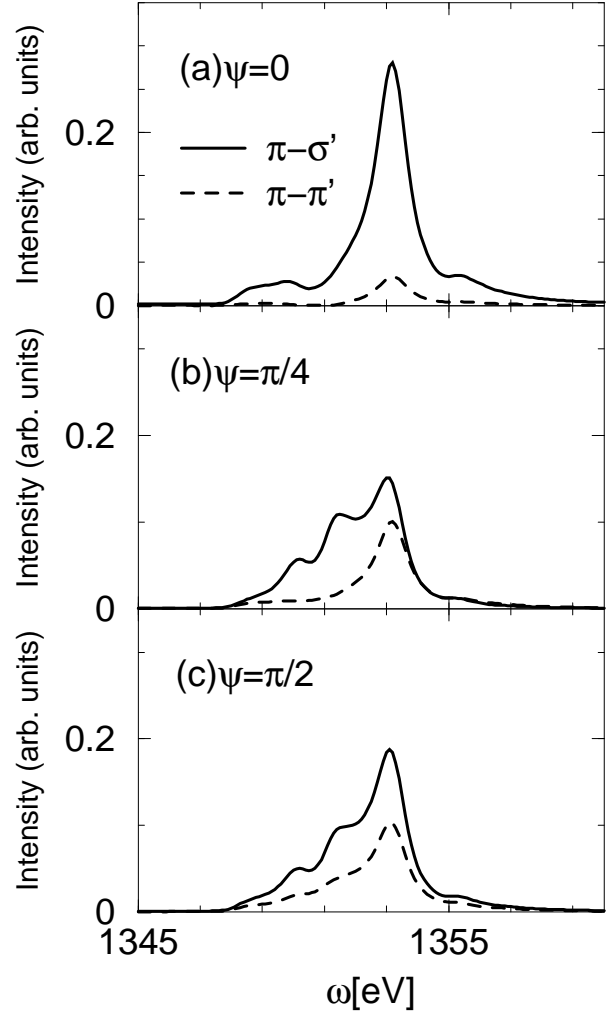}%
\caption{\label{fig.average}
The RXS spectra at the Ho $M_5$ edge with
$\textbf{G}_2=(0,0,\tau_2)$
in the spin slip phase with $N=11$.
The intensities are averaged between the anti-clockwise and
clockwise helical phases.
The solid and broken lines represent
the $\pi-\sigma'$ and $\pi-\pi'$ channels,
respectively.
The azimuthal angles are chosen as (a) $\psi=0$,
(b) $\psi=\frac{\pi}{4}$, and (c) $\psi=\frac{\pi}{2}$.
}
\end{center}
\end{figure}

We investigate the RXS spectra at the satellite spot
in the spin slip phase. The qualitative evaluation in 
Sec. \ref{sect.5.3} indicates that
they are the mixture of the profiles 
$\alpha_{\textrm{E1}}^{(1)}(\omega)$ and
$\alpha_{\textrm{E1}}^{(2)}(\omega)$.
The dominance of the former over the latter leads
to the expectation that the spectra at the first and
second satellite spots in this phase are governed
by the former.
The situation at the second satellite spot is
of particular interest since the the spectra consist
of the pure $\alpha_{\textrm{E1}}^{(2)}(\omega)$
both in the conical and the uniform helical phases.
Quantitatively, 
we calculate the RXS spectrum at the second satellite
spot in a spin slip phase with $N=11$.
The core hole lifetime broadening is set to be 0.6 eV.
As shown in Sec. \ref{sect.5.4}, the spectra
exhibit the domain dependence, that is,
the clockwise or anti-clockwise alternation of the
direction of the magnetic moment affect the spectra.
Thus we take the domain average of the spectra
assuming equal probability of the two domain.
Moreover, the spectrum varies as a function of
$\psi$ in this phase.
The results are summarize in Fig. \ref{fig.average}.
Although the spectral shape varies with $\psi$,
the entire shape including the highest peak position
is dominated by the pure 
$|\alpha_{\textrm{E1}}^{(1)}(\omega)|^2$
as anticipated qualitatively.

This fact brings us to the speculation of the
spectral shape evolution.
With decreasing temperature from $T_{\textrm{N}}$
down to $T_{\textrm{C}}$, the spectral shape at the
second satellite spot evolves from
the pure $|\alpha_{\textrm{E1}}^{(2)}(\omega)|^2$,
through the 
$|\alpha_{\textrm{E1}}^{(1)}(\omega)|^2$ dominated one,
into the original one
accompanying the shift of peak position from 1353.2 eV
to 1351.4 eV and again to 1353.2 eV, respectively,
large enough to detect experimentally.
An experimental observation
of the second satellite spot spectra in the different
phases is desirable.

\section{\label{sect.7}Concluding remarks}
Stimulated by the RMXS experiments,\cite{Spencer05,Ott06}
we have investigated the Ho $M_5$ spectra at the
first and second satellite spots.
The analysis have performed by exploiting the 
useful expression of the scattering amplitude
derived on the basis of the 
localized electron picture.\cite{Nagao05,Nagao06}
Although a large enhancement of the intensity
is expected at the Ho $M$ edge in the RXS spectrum,
the spectrum shows no azimuthal angle dependence
in the uniform helical and conical phases with the 
choice of $\textbf{G}_m$ adopted in the experiment.
Thus, we have mainly focused on the spectral shape analysis.

A qualitative analysis have revealed that
the RXS spectrum at the first satellite spot consists
of the pure rank one energy profile 
($\alpha_{\textrm{E1}}^{(1)}(\omega)$) in the
uniform helical phase.
Upon cooling temperature, the spectrum comprises
that of the rank two 
($\alpha_{\textrm{E1}}^{(2)}(\omega)$) as well as 
$\alpha_{\textrm{E1}}^{(1)}(\omega)$ both in the
spin slip and conical phases.
On the other hand, the spectrum at the second satellite spot
is made up of the pure $\alpha_{\textrm{E1}}^{(2)}(\omega)$
in the uniform helical and conical phases.
In the spin slip phase, however, the spectrum
includes $\alpha_{\textrm{E1}}^{(1)}(\omega)$ too.
Our numerical calculation have revealed that
the magnitude of the 
$|\alpha_{\textrm{E1}}^{(1)}(\omega)|$
is much larger
than $|\alpha_{\textrm{E1}}^{(2)}(\omega)|$, 
for instance,
$|\alpha_{\textrm{E1}}^{(1)}(\omega)|/
|\alpha_{\textrm{E1}}^{(2)}(\omega)| \simeq 27.7$
with $\Gamma=0.6$ eV.
Thus, whenever both profiles mix, 
we can expect the $\alpha_{\textrm{E1}}^{(1)}(\omega)$
dominates the spectral shape.
Then, the spectrum at the first satellite spot
looks like $|\alpha_{\textrm{E1}}^{(1)}(\omega)|^2$
at all magnetic phases.
Contrary to this,
we conclude
that the spectral shape of the second satellite
spot varies as a function of temperature.
That is, the spectrum is pure 
$|\alpha_{\textrm{E1}}^{(2)}(\omega)|^2$ in the helical
phase, then looks like 
$|\alpha_{\textrm{E1}}^{(1)}(\omega)|^2$ in the spin slip
phase, and finally, becomes pure 
$|\alpha_{\textrm{E1}}^{(2)}(\omega)|^2$ again 
in the conical phase.
If such variation of the spectral shape would be
observed by the experiment, it demonstrates
a clear evidence
of the change of the magnetic structures.

If the experiment in the conical phase is attainable,
another attractive outcome can be expected.
That is, the RXS spectrum from the higher-order harmonic
satellite spots may be detected.
It is brought about by the distortion of the orientation
of the magnetic moment within the basal plane.
In neutron scattering, the signals at the
$(6 \pm 1)$th magnetic satellite spots were
observed.\cite{Koehler66}
In the present case of RXS in the $E$1 transition,
the intensities at the $(6 \pm 2)$th as well as
$(6 \pm 1)$th order satellite spots, which are related
to rank two and one operators, respectively,
will be detected.
By evaluating the Ho case
in the vicinity of the $M_5$ absorption edge,
we can conclude that the intensity at the 
fourth satellite spot is within the reach of the
present experimental condition of the $E$1 transition
in the $\pi-\sigma'$ channel.

As for the numerical results, first, we
have concentrated on the comparison of our
result with the experiment in the uniform helical phase.
The calculated RXS spectrum at the first satellite spot
and the absorption coefficient show excellent agreement
with the observed ones both the spectral shape and the
peak position 
(for the former spectrum).\cite{Ott06,Thole85,Schille93}
The agreement assures the reliability of the
present analysis
based on the localized picture.
At the same time, it also demonstrates the effectiveness of the
spectral shape analysis in the RXS theory, which is 
rare, in particular, in the $f$-electron systems.
The agreement of the spectrum at the second satellite spot
between ours and the experiment is not an ideal one.
Although the peak position, below about 1.8 eV compared
to that of the first satellite spectrum, is properly
reproduced, the theory includes an extra peak 
around 1353.2 eV which is missing in the experiment. 
The reason of this discrepancy is still unclear.

Next, our evaluation of the spectra at the $M_4$ edge
showed that the magnitudes of them are roughly
two orders of magnitude smaller than those
at the $M_5$ edge.
Thus, the possibility of experimental 
detection seems to be slim
at the second satellite spot.
On the other hand, the magnitude of the
spectrum at the first satellite spot at the $M_4$ is expected
to be the same order as that of the second satellite
spot at the $M_5$ edge, and seems to be detectable.
Finally, we have confirmed that the spectrum
the second satellite spot in the spin slip phase
certainly governed by the 
$|\alpha_{\textrm{E1}}^{(1)}(\omega)|$ as anticipated.

A qualitative part of the present analysis may be applicable
to the interpretation of the RMXS experiment
near the Ho $L_3$ edge in the $E$1 process,
by neglecting the band nature of the $5d$ electrons as
the previous works had followed.\cite{Gibbs88,Hannon88,Gibbs91}
Since the absorption coefficient at the $L_3$ edge exhibited
no multi-peak structure, we conclude either the multiplet
splitting of the intermediate are not large enough and/or
the energy resolution were not fine enough
to distinguish the line shapes of the different profiles
$\alpha_{\textrm{E1}}^{(1)}(\omega)$ and 
$\alpha_{\textrm{E1}}^{(2)}(\omega)$.
Thus, only we can guess is the polarization analysis.
If the sample used were in the uniform helical phase,
eqs. (\ref{eq.helical.G1}) and (\ref{eq.helical.G2})
state that the spectra at the first and second 
satellite spots are pure $\alpha_{\textrm{E1}}^{(1)}(\omega)$ 
and $\alpha_{\textrm{E1}}^{(2)}(\omega)$, respectively.
The former is absent in the $\sigma-\sigma'$ channel.
The latter has contributions both in the
$\sigma-\sigma'$ and $\sigma-\pi'$ channels.
These conclusions are just the ones 
deduced previously.\cite{Gibbs88,Hannon88,Gibbs91,Hill96}
Note that, in ref. \citen{Gibbs91},
although the authors concluded there was no
$E$1 contribution in the $\sigma-\pi'$ channel
at the second satellite spot $(0,0,2+2\tau)$,
the data exhibits a small hump around 8071 eV, at which
the $E$1 peak is expected.
By substituting the value of the Bragg angle ($\theta^{(2)}
\simeq 18^{\circ}$) in 
eq. (\ref{eq.helical.G2}), 
the intensity of the $\sigma-\pi'$
channel in the $E$1 process is evaluated about a
tenth of that in the $\sigma-\sigma'$ channel. 
Since the ratio is nearly equal to that observed,
the data may show the contribution from the $E$1 process
in the $\sigma-\pi'$ channel at the second satellite spot.
Thus the intensity around this energy may show the resonant
behavior.  Additionally, the analysis on the $E$2 process
by means of the present theory qualitatively shows
complete agreement with the previous 
interpretation.\cite{Gibbs88,Hannon88,Hill96}
Because our formalism allows naturally inclusion of the
energy profile analysis, such an investigation is the
next step along this line.\cite{Com.E2}

Since the wavelength of soft x-ray is suitable
for the investigation of the long periodic magnetic
structure observed in several heavy rare earth material, 
the present theory may be effective for the analysis
of the R(M)XS spectra at the $M_{4,5}$ edges of such
heavy rare earth compounds, for example, DyB$_2$C$_2$.
Actually, the absorption coefficient in the vicinity
of the Dy $M_5$ edge exhibits multi-peak structure
similar to that observed in Ho.\cite{Thole85,Mulders06}
On the other hand, the RXS spectra differ significantly
between those of Dy and Ho.  
The spectrum of Dy extends over a few tens eV, an order 
of magnitude wider than that of Ho,
which may not be explained within the present
localized electron picture.\cite{Mulders06}
The work along this line is a future study.

\begin{acknowledgments}
We thank M. Takahashi for valuable discussions.
This work was partially supported by a Grant-in-Aid for Scientific Research 
from the Ministry of Education, Science, Sports and Culture, Japan.
\end{acknowledgments}

\appendix
\section{\label{app.A}The geometrical factors and 
the scattering amplitude}
Here, we present the geometrical factors needed to 
evaluate eqs. (\ref{eq.con.1}) and (\ref{eq.con.2}).
The results are classified by the photon polarizations.
The origin of $\psi$ is defined such that
the $y$ axis lies in the scattering plane.
The geometrical configuration we adopted
in this work can be found in our previous paper.\cite{Nagao06}
For the first satellite spot $\textbf{G}_1=
(0,0,\tau_1)$, 
\begin{equation}
(\mbox{\boldmath{$\epsilon$}}' \times
\mbox{\boldmath{$\epsilon$}})_x
-i (\mbox{\boldmath{$\epsilon$}}' \times
\mbox{\boldmath{$\epsilon$}})_y
=\textrm{e}^{i \psi} \cos \theta^{(1)}
\left\{
\begin{array}{lc}
0  & \textrm{for} \ \sigma-\sigma'\\
-i & \textrm{for} \ \sigma-\pi'\\
 i & \textrm{for} \ \pi-\sigma'\\
-2 \sin \theta^{(1)} & \textrm{for} \ \pi-\pi'\\
\end{array}
\right. , \label{eq.App.A1}
\end{equation}
and
\begin{equation}
P_{4}^{(2)}
(\mbox{\boldmath{$\epsilon$}}', \mbox{\boldmath{$\epsilon$}})
-i P_{3}^{(2)}
(\mbox{\boldmath{$\epsilon$}}', \mbox{\boldmath{$\epsilon$}})
=\frac{\sqrt{3}}{2} \textrm{e}^{i \psi} \cos \theta^{(1)}
\left\{
\begin{array}{lc}
0 & \textrm{for} \ \sigma-\sigma'\\
1 & \textrm{for} \ \sigma-\pi'\\
1 & \textrm{for} \ \pi-\sigma'\\
0 & \textrm{for} \ \pi-\pi'\\
\end{array}
\right. ,
\end{equation}
where $\theta^{(1)}$ is the Bragg angle associated with
$\textbf{G}_1$. Then, for the second satellite spot
$\textbf{G}_2$,
\begin{equation}
P_{1}^{(2)}
(\mbox{\boldmath{$\epsilon$}}', \mbox{\boldmath{$\epsilon$}})
-i P_{5}^{(2)}
(\mbox{\boldmath{$\epsilon$}}', \mbox{\boldmath{$\epsilon$}})
=\frac{\sqrt{3}}{2} \textrm{e}^{2 i \psi}
\left\{
\begin{array}{lc}
1 & \textrm{for} \ \sigma-\sigma'\\
 i \sin \theta^{(2)} & \textrm{for} \ \sigma-\pi'\\
-i \sin \theta^{(2)} & \textrm{for} \ \pi-\sigma'\\
\sin^2 \theta^{(2)} & \textrm{for} \ \pi-\pi'\\
\end{array}
\right. , \label{eq.App.A3}
\end{equation}
where $\theta^{(2)}$ is the Bragg angle associated with
$\textbf{G}_2$. 
Note that a relation
$\sin \theta^{(2)}=2 \sin \theta^{(1)}$ holds.

By substituting these results into eqs. (\ref{eq.con.1})
and (\ref{eq.con.2}), we obtain the RXS intensities
at the first satellite spot as
\begin{eqnarray}
|f_{\textrm{1}}|^2 &\propto&
\left[\frac{N}{2}\cos\left(\gamma-\frac{\pi}{N}\right) J 
\sin \beta \cos \theta^{(1)} \right]^2
\nonumber \\
&\times& 
\left\{
\begin{array}{lc}
0 & \textrm{for} \ \sigma-\sigma'\\
\left| \alpha_{\textrm{E1}}^{(1)}(\omega) 
+ \frac{3(2J-1)\cos\beta}{4}
\alpha_{\textrm{E1}}^{(2)}(\omega) \right|^2
 & \textrm{for} \ \sigma-\pi'\\
\left| \alpha_{\textrm{E1}}^{(1)}(\omega) 
- \frac{3(2J-1)\cos\beta}{4}
\alpha_{\textrm{E1}}^{(2)}(\omega) \right|^2
 & \textrm{for} \ \pi-\sigma'\\
4 \sin^2 \theta^{(1)} 
|\alpha_{\textrm{E1}}^{(1)}(\omega) |^2
 & \textrm{for} \ \pi-\pi'\\
\end{array}
\right. , \nonumber \\
\label{eq.App.A4}
\end{eqnarray}
and the one at the second satellite spot as
\begin{eqnarray}
|f_{\textrm{1}}|^2 &\propto&
\left[\frac{3N}{16}\cos\left(2\gamma-\frac{2\pi}{N}\right) 
J(2J-1) \sin^2 \beta  \right]^2
\nonumber \\
&\times& |\alpha_{\textrm{E1}}^{(2)}(\omega) |^2
\left\{
\begin{array}{lc}
1 & \textrm{for} \ \sigma-\sigma'\\
\sin^2 \theta^{(2)} & \textrm{for} \ \sigma-\pi'\\
\sin^2 \theta^{(2)} & \textrm{for} \ \pi-\sigma'\\
\sin^4 \theta^{(2)} & \textrm{for} \ \pi-\pi'\\
\end{array}
\right. . \label{eq.App.A5}
\end{eqnarray}
These results show the spectral intensities
exhibit no azimuthal angle dependence.


\begin{thebibliography}{99}
\bibitem{ComScreen}
        In solids, the magnitude of the Slater integral
        is reduced because of large screening effect.
        The isotropic and the anisotropic parts of the 
        Slater integral are reduced their values
        by multiplying factors 0.25 and 0.8, respectively.
\bibitem{Com2}
          Data reported in ref. \citen{Spencer05}
          were measured for bulk sample.
          However, the authors themselves
          said the precision of the
          spectral shapes of their data was suspicious.
          We do not compare our results with them
          in the present work.
\bibitem{Com1}
          Since the experimental data in ref. \citen{Ott06}
          are given as the scattering amplitudes, we plotted
          them after taking the square of the original data.
\bibitem{Com.E2}
          When we discuss the $E$2 process, the geometrical 
          factors appeared in ref. \citen{Nagao06} 
          should be replaced by the corrected ones.
          eqs. (2.18), (2.19), (B1)$\sim$(B5), (B7), and (B8)
          should be multiplied by $3/4$, while eq. (B6) by 
          $\sqrt{3}/4$.
          The coefficient of $P_{\mu}^{(1)}$ in eq. (B2) and
          that of $P_1^{(0)}$ in eq. (B4) should be replaced
          by $1/2$ and $-\sqrt{5}$, respectively.
\end{thebibliography}

\end{document}